**ORIGINAL PAPER**

# Ethics of generative AI and manipulation: a design-oriented research agenda

Michael Klenk[1]

## Abstract
Generative AI enables automated, effective manipulation at scale. Despite the growing general ethical discussion around generative AI, the specific manipulation risks remain inadequately investigated. This article outlines essential inquiries encompassing conceptual, empirical, and design dimensions of manipulation, pivotal for comprehending and curbing manipulation risks. By highlighting these questions, the article underscores the necessity of an appropriate conceptualisation of manipulation to ensure the responsible development of Generative AI technologies.

## Introduction

Research on generative AI is growing at scale, and the results achieved by recent applications are nothing short of astonishing (though see Floridi, 2023). These developments create "enormous promise and peril" (The Economist, 2023), especially by enabling effective, automated influence at scale.

On the one hand, such ability is promising because many good things depend on effective influence. For example, effective influence is required to facilitate better lifestyle interventions to improve health outcomes (see e.g. Tremblay et al., 2010). It could also improve public policy, helping governments to communicate with citizens amidst the noise of propaganda, filter bubbles, and fake news (European Commission, forthcoming).

On the other hand, effective influence invites manipulation, a morally dubious form of influence. Generative AI could, for instance, "make email scams more effective by generating personalised and compelling text at scale" (Weidinger et al., 2022) or learn to generate outputs that effectively exploit users' cognitive biases to influence their behaviour (Kenton et al., 2021). More generally, whenever effective influence is rewarded—which is the case in almost any area of human interaction, such as social life, marketing, or politics—there is a strong incentive to turn from legitimate forms of influence like rational persuasion to more effective but morally dubious forms of influence like manipulation. Hence, generative AI "aggravates" (Klenk & Jongepier, 2022b) existing ethical concerns about online manipulation.

However, there is no clear view of how the (dis-)value of manipulation should play a role in designing new technologies based on generative AI. How, in other words, can generative AI (or, more precisely, applications that use it) be designed so that its application avoids illegitimate forms of manipulation? Existing work in AI ethics barely touches on design questions and focuses more on the important but still preliminary step of drawing attention to pertinent ethical risks (e.g. Weidinger et al., 2022). Moreover, some technical work on AI alignment already addresses in general terms how to make generative AI applications "helpful, honest, and harmless" (Askell et al., 2021), but there is insufficient attention paid to an appropriate conceptualisation of manipulation that can guide design, which is unsurprising given that manipulation is a difficult concept to grasp.

The lack of attention to manipulation in the debate about generative AI is a significant omission. Manipulation is identified as a disvalue and thus an explicit target of AI regulation, e.g. in the EU's forthcoming AI Act (European Commission, 2021; European Commission et al., 2022). More

✉ Michael Klenk
 M.B.O.T.Klenk@tudelft.nl

1  Department of Values, Technology and Innovation, TU Delft, Jaffalaan 5, 2628 BX Delft, The Netherlands

generally, manipulation is considered a threat to democracy and trustworthiness, which means that it is a fundamental threat to the critical aim of responsible, trustworthy AI (Faraoni, 2023). In addition, a large body of literature documents worries about manipulation in other contexts, notably nudging and advertising (cf. Sunstein, 2016). There is, thus, a compelling legal and moral case for paying attention to manipulation. Given these goals, it is imperative to understand clearly what manipulation is and to devise appropriate requirement specifications.

Therefore, this article discusses a research agenda studying manipulation in generative AI. I argue that good research on manipulation and generative AI—which everyone concerned with trustworthy AI and the value of democracy is or should be interested in—depends significantly on our conceptualisation of manipulation. It matters because different phenomena will come into view depending on our conception of manipulation. It also matters pragmatically because different conceptions of manipulation will imply different design and regulatory requirements.

I proceed as follows. The section "Design for values and conceptual engineering" "Design for non-manipulation" introduces the design for value approach in general. The section "Design for non-manipulation" then discusses pertinent research questions about manipulation that relate to the conceptual, empirical, and implementation phases of a design for value project, with a focus on the conceptual phase.

## Design for values and conceptual engineering

I take a design perspective that aims to help designers and engineers put values at the heart of the design of new technologies (van de Poel, 2020; van den Hoven et al., 2015). Central to the design perspective—whose importance is stressed by the IEEE, the WHO, UNESCO, the EU, and many others—is that human values should inform and shape appropriate design requirements.[1] Consequently, several key questions for any design for value project concern the nature of the values that should be designed for.[2]

Central to the idea of design for values is then that the target values can be specified in a way that allows for a systematic and reliable deduction of concrete design requirements from a general, abstract conception of target values such as 'trust,' 'democracy,' or 'non-manipulation' (van de Poel, 2013, 2020; Veluwenkamp & van den Hoven, 2023). It is generally acknowledged that there are often different, (prima facie) plausible conceptualisations of target values, and quite some attention has been devoted to different ways of settling on a value conceptualisation (cf. Friedman & Hendry, 2019).[3]

However, what's only recently been emphasised is the important question of how we can adjudicate between different, perhaps conflicting conceptualisations of a target value (Himmelreich & Köhler, 2022; Veluwenkamp & van den Hoven, 2023).[4] As Veluwenkamp and van den Hoven (2023, p. 2) put it, "it is not always obvious which concepts invoked in the decomposition of requirements is the most appropriate in the relevant context of use." In answering the question of *how can we decide which conceptualisations to use?*, we must be aware that conceptualisations have consequences; they matter a great deal. For one, different conceptualisations matter for our understanding because they will bring different phenomena into view. For example, conceiving manipulation as an influence hidden from the user will prompt researchers and designers to look at completely different phenomena than conceiving manipulation as a kind of social pressure that need not be hidden from the user at all. In that sense, different conceptions of manipulation function like searchlights. Once they are adopted for a given target value, they bring into scope some and blind us to other phenomena, which may be equally if not more important

---

[1] See, for example, European Parliamentary Research Services (2020), IEEE (2019). In line with the relevant literature, I am using the term 'value' quite broadly here to mean something like 'a phenomenon of positive normative significance.' In that sense manipulation is not a value but a dis-value, a phenomenon of negative normative significance. This loose way of talking seems appropriate in this context, and it should not be interpreted as leaning on more nuanced, axiological discussions.

[2] A preliminary question for any design for value project concerns the kind of values that should be designed for, i.e. an enumeration of the target values. The answer to this question is not—in general—trivial or obvious. Which values matter in which context is a complex

Footnote 2 (continued)
ethical and societal question. But in our case, this question has in part been answered by the moral and legal case for attending to manipulation, to which I already pointed in the introduction. Naturally, this does not mean, however, that manipulation should be the exclusive or even dominant focus in pursuing responsible generative AI. See Weidinger et al. (2022) for a taxonomy of other risks, many of which are perfectly general worries about AI, such as worries about privacy, fairness, and explainability.

[3] In what follows, I use 'conceptualisation' and 'conception' interchangeably. I use 'conceptualisation' rather than 'concept' to emphasise the sense in which we (artificially) construct conceptualisations, e.g. for scientific use, and to demarcate the discussion from concepts as the building blocks of thought.

[4] A conceptualisation or conception of a concept can be thought of as a specification or description of a concept's content. Concepts have a content and an extension. A concept's content can be thought of as its specification or a description of what the concept is about. A concept's extension, in contrast, refers to the things that the concept is about. For example, the content of the concept 'bachelor' is something like 'an unmarried man' (the concept is 'about' unmarried men), while its extension contains all unmarried men. While some questions are about content—what is the concept about—others concern extension.

(see also Barnhill, 2022). Therefore, it is relevant for good research on manipulation and generative AI that the chosen conceptualisation reflects or covers the phenomena that make people worried about manipulation in the first place.

Furthermore, conceptualisations also influence the concrete technological interventions and innovations developed to solve the design challenge. For example, thinking of *trust* as *epistemic reliability* will imply very different design requirements, and result in different technical solutions toward the goal of *trustworthy AI* than conceptualising *trust* in *moral terms* such as *benevolence* (cf. Veluwenkamp & van den Hoven, 2023). Picking a conceptualisation is thus far from being 'just about words.' It is a consequential, material choice. When we start with two different conceptualisations of manipulation, we will likely get two different technical artefacts or systems when we design for non-manipulation. Moreover, if our conceptualisation is bad or inappropriate, the design challenge addresses a faux problem. So, when we aim to design for values, our success depends on the kinds of conceptualisations we pick.

Therefore, good research on manipulation and generative AI depends on an appropriate conceptualisation of 'manipulation.'[5] Existing discussions of manipulation and generative AI leave much wanting in that dimension. Weidinger et al. (2022) are concerned with a taxonomy of generative AI risks. When they discuss manipulation, they fail to distinguish it from deception adequately. This omission raises questions that they do not answer. Is design for non-manipulation just design for non-deception? Or is there more? If there is more, what would that conception look like? Kenton et al. (2021) provide a more elaborate discussion, and they end up with a broad and encompassing conceptualisation of manipulation, arguing from a safety perspective: the more phenomena covered, the safer the resulting design. But, as they acknowledge themselves, their conceptualisation may be "too wide-ranging" (Kenton et al., 2021, p. 11). Too many phenomena will come into view as instances of manipulation, cloud our sense of what manipulation really is, and designs targeted at the phenomena may be overburdened with requirements. Going forward, research on manipulation in generative AI should focus on sharper, more appropriate conceptions of the target phenomenon.

The obvious yet fundamental question concerns the appropriate criteria for choosing a conceptualisation. What makes one conception of, for example, 'manipulation' better than another? Traditionally, conceptualisations seem appropriate insofar as they match the target phenomenon. In that view, a conceptualisation of manipulation is appropriate insofar as it captures all and only cases of manipulation. Let this be the *narrow* criterion of *appropriateness*.[6] Importantly, a conceptualisation of manipulation is appropriate according to the narrow criterion quite independently of whether it 'works' in practice, such as in design or policy work. The narrow criterion chiefly aims at understanding by clarifying the constituent parts of a concept with little to no regard for whether or not the conceptualisation is helpful in design projects.

However, recently, the debate on 'conceptual engineering' in philosophy and the ethics of technology suggested that there may also be moral and pragmatic reasons that have a legitimate influence on our choice of conceptualisation, and tentative proposals have been made about how to systematically assess those reasons (cf. Veluwenkamp & van den Hoven, 2023). From this perspective, moral and pragmatic considerations about the causal effects of using a particular conceptualisation or its practicality *also* play a role in determining whether it is an appropriate conceptualisation (in addition to considerations about whether the conceptualisation captures all and only cases of the target phenomenon, in line with the narrow criterion). Let this be the *broad* criterion of *appropriateness* for conceptualisation choice. The broad criterion of appropriateness may especially be relevant from a design perspective, given that a conceptualisation of manipulation in the context of generative AI ultimately ought to inform design choices. However, to what extent broad considerations ought to outweigh narrow considerations is a challenging and unresolved metaphilosophical question.

My aim here is not to weigh in on the metaphilosophical question of whether and why we should prefer the narrow or broad criterion of appropriateness.[7] Instead, in what follows, I will point out the open questions that still stand in the way of contributing to either approach: What *is* manipulation (as a folk concept), and what should it be, provided we are prepared to deviate from the folk concept, for reasons of accuracy or other pragmatic, and moral reasons?

---

[5] Implicitly, this point seems to be acknowledged, for example, in much of the (applied) ethical debate on manipulation in current online technologies Klenk and Jongepier (2022a) or nudging Wilkinson (2013), where researchers often first aim to arrive at an appropriate conceptualisation of manipulation before commencing to analyse specific cases through that lens. This mirrors a kind of mid-level principle approach to bioethics, cf. Flynn (2022).

[6] This view is closely linked to the method of conceptual analysis in philosophy, see Klenk and Jongepier (2022b, pp. 16–19) for discussion.

[7] This is still a controversially debated issue in the debate about conceptual engineering and conceptual ethics. Though tentative proposals have been made, I am sceptical that anything approaching a theory of conceptualisation choice is currently available. A critical open question is when we ought to consider a given conceptualisation defective.

## Design for non-manipulation

Design for value approaches typically involve the following stages: a phase of considering the appropriate conceptualisation of a value using *conceptual* means (e.g. reasoning), an empirical stage where stakeholder input is solicited to contribute to the conceptualisation, and a design or implementation stage (Buijsman et al., forthcoming; Friedman & Hendry, 2019).

The three stages of a design for value project—conceptual, empirical, and design—are meant to be repeated at different stages of specification of the target value (i.e. from identification of the value to conceptualisation, association with norms, etc.), until concrete design requirements are reached (cf. Veluwenkamp & van den Hoven, 2023). I restrict my focus primarily to the conceptualisation stage. As our understanding of manipulation grows and questions about the appropriate conceptualisation get resolved, we should expect the debate to turn to the subsequent steps of operationalising toward concrete design requirements.[8]

Since manipulation is generally seen as a dis-value, I focus on *non-manipulation*, viz. the absence of manipulation, as a target value. It is clear, then, that even a successful non-manipulative design will probably still leave many other ethically significant issues untouched. A generative AI application that does not manipulate may be ethically legitimate *from a manipulation perspective*, but *overall* it may still have other ethical issues (such as issues to do with explainability, privacy, etc.). As such, design for non-manipulation may need to be combined with, or form a part of, broader design aspirations, such as design for trustworthy AI or design for democracy (EGE, 2023).

## Conceptual stage

To design for non-manipulative generative AI, at least the following questions need to be answered:

1. What are reliable criteria to identify manipulation and to distinguish it from other (often less morally suspect) forms of influence?
2. How can generative AI applications be aligned with criteria for non-manipulation?
3. When and why is manipulation morally bad?

The first question is quintessentially connected to an appropriate conceptualisation of manipulation. Answering it will give us a way to tell whether a given influence—such as an output produced by a generative AI application—is manipulation. For example, suppose that a personal digital health assistant driven by generative AI outputs 'You should be ashamed of yourself for ordering that meal' to the user after drawing on their recent purchase history. To decide whether that prompt—or any other output generated by the system—qualifies as manipulation, we need reliable criteria to identify manipulation. In this section, I will briefly review the most pertinent criteria for manipulation. After considering and rejecting several potential criteria, I will suggest—in Sect. "The indifference criterion"—that the indifference criterion is most appropriate to conceptualise manipulation.

### The continuum model of influence

Manipulation is a form of influence (Coons & Weber, 2014b). As social animals, humans influence each other in countless ways. Some influences are intentional, such as a speech act, while others are unintentional, such as the intimidating effect a very tall person may have on others. However, not all intentional influences are ethically problematic. For example, if you are the passenger in a car and you yell out to the driver to warn them about an accident, you are not doing anything wrong (cf. Sunstein, 2016). Therefore, the first question requires us to determine how manipulation, as a morally suspect influence, is set apart from other types of influence that are generally deemed legitimate.

Criteria for identifying manipulation implied by a chosen conceptualisation may be derived by contrasting it with other forms of influence. Indeed, it has been suggested that manipulation sits on a continuum of influence, situated between rational persuasion and coercion (Beauchamp, 1984; Beauchamp & Childress, 2019). This continuum model helps draw basic distinctions and conceptualise the idea that there are some benign types of influence, like rational persuasion, and other types of influence that are clearly problematic, like coercion.

---

[8] A design focus implies at least one important assumption and limitation. It assumes that there are people motivated to design for non-manipulation. I need not assume that people are morally motivated, however. Existing and forthcoming regulation on manipulation should provide some purely pragmatic impetus to seek ways to design non-manipulative generative AI. What this leaves out is the question of how to regulate or control for non-manipulative AI (both of which can be approached from a design perspective, of course). Note also that I will not discuss how do balance the goal of non-manipulation with other values. An important aspect of design approaches to values in technology is that they will have to deal with conflicting values van de Poel (2015). For instance, the design of an engine will strike a balance between cost-effectiveness and sustainability or content moderation at a social media platform realises values of the decision makers. Applications based on generative AI will likewise have to strike a legitimate balance between the promise of effective influence and the peril of manipulation. There will be many other trade-offs and conflicts that a full-blown design for value approach to generative AI will also have to consider (e.g. concerning sustainability and resource-use). The focus of this research agenda, however, will be firmly on questions about manipulation, thus leaving open the further question of balancing concerns about manipulation appropriately with other goals and values.

However, the continuum model does not yet provide us with reliable criteria for manipulation. There seem to be forms of non-persuasive and non-coercive influence that are not manipulation (Noggle, 1996). For example, dressing up for a job interview is neither rational persuasion nor coercion, but it does not look like manipulation either (Noggle, 1996). Depending on how we define the reference points of 'persuasion' and 'coercion,' the continuum model might give us criteria for manipulation that are much too broad, resulting in overly stringent design requirements for generative AI.

Therefore, it is more promising to turn to philosophical theories of manipulation that offer more specific criteria for identifying manipulation. There are several influential ideas about manipulation that are simple, intuitive, and seemingly easy to apply in practice.

**The hidden influence criterion**

Perhaps the most influential idea is that manipulation is necessarily a form of hidden influence (cf. Faraoni, 2023, and its uptake and reflection in policy documents). According to Susser et al. (2019a, 2019b), manipulation is an influence that the victim is not or could not easily be aware of. For this conception to be useful in generative AI, it is crucial to specify exactly what remains hidden from the manipulation victim. For example, must the intended outcome of the influence be hidden from the user? Or the precise psychological mechanism through which the influence is intended to work? Or how the influence was generated? The latter, for example, would suggest that any influence generated by generative AI but not declared as such would count as manipulative on the hidden influence conception. In any case, the hidden influence conception helps distinguish manipulation from persuasion and coercion on the continuum model because these forms of influence are necessarily overt (cf. Klenk, 2021c).[9]

However, the hidden influence conceptualisation of manipulation is unlikely to provide reliable criteria to capture the phenomenon of manipulation accurately, let alone entirely.

On the one hand, many hidden influences do not fall under manipulation. For instance, the heuristic and biases research programme in psychology suggests that many of our decisions arise out of hidden processes that are not the result of conscious deliberation (Kahneman, 2012). Still, such processes often seem legitimate and non-manipulative (cf. Sunstein, 2016). Therefore, the criterion of hidden influence risks being over-inclusive: it classifies too many cases as manipulation, thus generating false positives. It would require further work to explain *how* hidden influence is to be understood in a way that makes it a credible criterion for manipulation.[10]

On the other hand, some important forms of manipulation are not covered by the hidden influence conception (cf. Klenk, 2021c). For example, a manipulative real-estate agent may use the homely scent of freshly baked cookies at a house viewing to lure in potential buyers who, nonetheless, are be fully aware that they are being manipulated (Barnhill, 2014). Similarly, the dark pattern known as a 'roach motel' often prevents users from cancelling a service by making it cumbersome and tiring to complete (Brignull, 2023). Victims of a roach motel are being manipulated even though they are often fully aware of the influence. Therefore, the hidden influence criterion also risks being under-inclusive: it generates insufficient cases as manipulation, thus generating false negatives.

As a result, the hidden influence conception fails given the *narrow* criterion of appropriateness I discussed in Sect. "Design for values and conceptual engineering" (recall that the narrow criterion says that a criterion is appropriate only if it captures all cases of manipulation).

It is also questionable whether the hidden influence conception fares well on a broad criterion of appropriateness. Setting aside the important questions raised at the beginning of this section, the criterion seems easy enough to apply, which may count in its favour given a broad criterion (though see Klenk, 2023). However, it may have the morally problematic implication that it shifts some of the burden for combating manipulation from the perpetrator to the victim (cf. Klenk, 2021c). After all, if manipulation is defined as hidden, then drawing it out into the open means that manipulation ceases to exist. This invites a simple but inappropriate approach to combating manipulation: by calling for potential victims of manipulation to sharpen their ability to uncover manipulation when a more appropriate approach would focus on regulating the perpetrator's behaviour instead. Thus, even if the over- and under-inclusiveness of the hidden influence conception could be addressed, there are moral reasons to think differently about the conceptualisation of manipulation, according to the broad criterion of appropriateness.

**The bypassing rationality criterion**

Another influential idea is that manipulation can be identified by influences that bypass rationality (Sunstein, 2016; Wilkinson, 2013). Again, the notion of *bypassing rationality* must be specified further for the criterion to be useful (see Gorin, 2014a for discussion). Like the hidden influence conception, the bypassing rationality conception should help to distinguish manipulation from coercion and persuasion, and it correlates with many paradigmatic cases of manipulation. For example, it is manipulative to prompt a generative AI to guilt-trip a target into donating money to a charity because the influence targets the victim's emotions and bypasses rational deliberation.[11]

---

[9] The hidden influence criterion may also be attractive because it lays a connection to the burgeoning debate about AI deception. However, it is crucial to recognise that manipulation and deception are *not* the same thing see Cohen (2023) for a recent assessment.

[10] Hidden influence may not even provide a necessary criterion for manipulation, as many researchers have pointed out, see Klenk (2021c).

However, important questions about the 'bypassing rationality' conceptualisation remain. While it seems accurate enough—it accounts for many paradigmatic cases of manipulation—it has been subject to severe criticism for generating false negatives (Gorin, 2014a, 2014b). Some forms of manipulation—such as peer pressure or charm—do not seem to involve bypassed rationality (Baron, 2003; Noggle, 2022). Hence, the bypassing conceptualisation of manipulation does not reliably identify all manipulation cases.

Moreover, many forms of tremendously important influences, such as testimony or influences that 'activate heuristics', bypass rationality but are not examples of manipulation. Hence, the bypassing criterion is also over-inclusive and generative false positives. For example, testimony bypasses rationality because it is often accepted at face value, given a positive evaluation of the source's credibility. This is not a rational process in the sense of being conscious, yet testimony is unlikely to be a form of manipulation. Similarly, the availability or recognition heuristic allows people to make frugal decisions without conscious deliberation. It is rational to rely on the heuristic when there is a correlation between the criterion and recognition (Gigerenzer & Goldstein, 1996). This suggests that 'activating' the availability heuristic need not be manipulative, even though it means to bypass rationality in the sense of bypassing conscious deliberation.

In summary, the bypassing rationality criterion suffers from over- and under-inclusivity. Since it also lacks the advantage of being relatively simple—insofar as bypassing rationality is more difficult to operationalise than hidden influence—it is of questionable relevance for the aim to design for non-manipulation.

### Disjunctive conceptions of manipulation

The hidden influence and bypassing conceptions fail because manipulation is a varied and diverse phenomenon. Neither the hidden influence nor the bypassing rationality conceptions offers a way to capture *all* cases of manipulation and *only* cases of manipulation.

This led some to wonder whether there is any conceptualisation of manipulation at all that is satisfactory given a narrow criterion for appropriateness (cf. Coons & Weber, 2014a; Klenk & Jongepier, 2022b).

However, disjunctive conceptions for identifying manipulation may be a solution. For example, in their discussion of the ethical alignment of language agents, Kenton et al. (2021) reflect on the diversity of philosophical accounts of manipulation and opt for a disjunctive conception that combines several criteria that are discussed in the philosophical literature. Accordingly, they suggest that manipulation occurs by bypassing rationality, trickery, or pressure.[12] Recent work on manipulation in AI ethics reflects a similar broad-strokes approach by throwing together different criteria like 'being hidden', which correlate with many cases of manipulation, hoping to capture the phenomenon in the wide net of a disjunctive conceptualisation.

However, disjunctive conceptualisations of manipulation are problematic from a narrow criterion of appropriateness (see Noggle, 2020, 2022). If a disjunctive conception incorporates criteria for manipulation that are over-inclusive on their own, then the resulting disjunctive conception risks being over-inclusive, too, viz. it wrongly classifies cases as manipulative. For example, including 'hidden influence' in a disjunctive conception risks inheriting the hidden influence criterion's problems with false positives. The worry that stems from a narrow conception of appropriateness may be addressed by interpreting the disjunction as tracking a family resemblance, such that individual disjuncts are not taken as sufficient for classification.[13]

However, disjunctive criteria still come with significant theoretical, practical, and ethical costs even if they address the problem of over-inclusivity.[14] Theoretically, they prevent us from identifying what the varied forms of manipulation have in common, for it is possible that there are simply different types of manipulation (cf. Coons & Weber, 2014a; Noggle, 2022). This is particularly worrisome given a narrow criterion of appropriateness. From a design perspective, we would need to specify what type of manipulation we are designing against each time. This is a practical problem independent of our criterion of appropriateness. A measure that may work against manipulation understood as hidden influence (e.g. disclaimers) may fail to address manipulation tracked by other disjuncts, like bypassing reason, yet all forms will register as 'manipulation' on a disjunctive conception. To remedy this, 'design for non-manipulation' could be misleading given a disjunctive criterion, and it would always have to specify exactly what kind of manipulation is in scope. This illustrates that there are definite practical advantages to identifying a common factor behind all forms of manipulation because it would make 'design for non-manipulation' clear and informative.

---

[11] For simplicity, I interpret 'bypassing rationality' in the psychological sense of 'bypassing conscious deliberation' following common understanding. Appealing to emotions is a paradigmatic example of doing this. There are more elaborate interpretations, discussed by Gorin (2014a), that suffer from similar problems to the ones I discuss here.

[12] Pressure is another criterion of manipulation that is sometimes discussed in the philosophical literature, cf. Noggle (2022).

[13] Thanks to an anonymous referee for suggesting this point.

[14] Some of these are problems arise given a narrow criterion of appropriateness. Though they do not undermine a positive evaluation of disjunctive criteria in general, or text-classifiers as a practicable way to implement disjunctive criteria, they matter for the appropriateness of a conceptualisation. Thanks to an anonymous referee for prompting me to clarify this point.

Ethically, disjunctive criteria make a common, unified ethical and regulatory response to manipulative influence more difficult (see Coons & Weber, 2014a for discussion). If there are different reasons why a given influence qualifies as manipulation, there have to be different ethical responses to it (a phenomenon known as supervenience). This is more complicated and stands in stark contrast to the current way regulators and ethicists propose to deal with manipulation—namely, in a uniform fashion. Therefore, insofar as an appropriate conceptualisation helps us understand and grasp the phenomenon in question, a disjunctive criterion merely dilutes the picture. This is clearly a problem given the narrow criterion for appropriateness.

On a broad criterion of appropriateness, disjunctive conceptualisations of manipulation fare better. There are already practicable disjunctive conceptualisations of concepts other than manipulation in the AI ethics domain. For example, text classifiers of hate speech can be understood as 'tracking' a disjunctive criterion for hate speech, and a similar criterion may be envisioned for manipulation.[15] Ultimately, however, there are serious obstacles to a disjunctive conceptualisation. A text classifier of manipulation would likely have to take into account a host of contextual factors that are hard to identify and represent. More so, there is likely no inherent connection between objectively identifiable features of the influence, such as the kind of words used in a text output and its manipulativeness. Manipulative influence does not, as it were, 'wear its manipulativeness on the sleeve.' For example, the sentence 'you promised to give it to me!' may be part of a manipulative guilt trip or part of a perfectly benign and non-manipulative conversation. It seems unlikely that we can reliably classify the influence without considering the motivation or genesis of the influence, such as the intention of the manipulator. This is because it is misleading to suggest, as Eliot (2023) does, that there are objectively identifiable manipulative patterns in texts that generative AI reproduces and that we could identify by looking at the generative AI output.[16] A text classifier as a practicable way to implement a disjunctive conceptualisation of manipulation would thus need to look at several currently unknown factors whose complexity needs to be considered in the evaluation of the approach.

In summary, disjunctive conceptualisations of manipulation are interesting but ultimately problematic on both narrow and broad criteria of appropriateness.

**The trickery criterion**

A more promising approach is to understand manipulation in terms of the influencer's intentions rather than the features of the influence itself. One very influential account suggests that we can identify manipulation by the intention to trick the recipient by causing them to violate a norm of belief, desire, or emotion (Noggle, 2020). Typical cases of fraud, for example, are classified as manipulation in this model because they involve the attempt to trick the target into adopting a false belief or an inappropriate desire. For example, when a scammer uses text messages to pose as a relative and asks for money, they try to induce a mistaken belief in the target.

The trickery conceptualisation seems helpful in addressing many intentionally manipulative uses of generative AI. In particular, the trickery conception works well in cases where generative AI is used as a tool to facilitate manipulative influence. In their critical assessment of AI-driven influence operations, Goldstein et al. (2023) describe how generative AI can be used to scale up fraud and make it more economical. For example, phishing and other attempts to make people solicit information or resources can aggravated by using generative AI to create persuasive phishing material, such as text messages or emails. The intent to trick the victim is clearly recognisable in such cases.

However, it is important to distinguish a different type of manipulation enabled by generative AI where the trickery criterion seems less appropriate.[17] In particular, the trickery conceptualisation produces false negatives in at least two relevant, though still less prevalent, use cases.[18]

---

[15] Thanks to an anonymous referee for suggesting this helpful example.

[16] This is explained, for example, by the failure of the bypassing rationality criterion. Since bypassing rationality understood as appealing to emotion is neither necessary nor sufficient for manipulation, it is unlikely that 'emotional' words or text patterns are reliable indicators of manipulative influence.

[17] Thanks to an anonymous referee for prompting me to clarify the non-intentional use-case.

[18] Noggle (2020) revises the account to focus on the intention to induce a *mistake* in the victim, mainly to accommodate a problem with false negatives in the trickery account regarding cases of pressure manipulation. I'd like to draw attention to two different types of cases that might lead to false negatives, and in these cases the observations about the trickery account apply to the revised mistake

First, someone may unwittingly use generative AI to generate manipulative influences, although they cannot be said to *intend* to trick anyone. For example, Brignull (2023) describes how automated A/B testing allows users to run the test *and* automatically implement the 'winning' design. Someone using this feature may simply be interested in creating an effective design that drives sales or engagement on their website. Still, since the 'winning' design may include paradigmatic dark patterns like, the user may be said to act manipulatively on account of their indifference or carelessness about the actual quality of their influence. The trickery account does not readily account for cases of unintended manipulation like this.[19]

Second, the trickery account's focus on intentions leads to problems insofar as generative AI—rather than being used as a tool for manipulation—may itself be manipulative. AI systems are generally thought to lack intention, even if the debate about this has been renewed in light of advances in generative AI. We might speak *as if* generative AI manipulates (Nyholm, 2022), and we might apply the criterion to analyse whether the deployers or designers of a generative AI system wanted to manipulate. But when the system itself is thought to be the source of manipulation, with no intention or only opaque (quasi-)intention, then the trickery account will yield a false negative: it will not classify such cases as manipulation even though they seem like cases of manipulation.[20] Cappuccio et al. (2022) argue that new forms of manipulation driven by AI may be "emergent" and not reducible to e.g. the intentions of a human user. Pham et al. (2022) also stress the importance of considering emergent, non-intentional forms of manipulation that have their source in the automated behaviour of AI-driven applications. An account of manipulation that emphasises the intention to trick or lead astray will not allow us to identify unwitting manipulation that emerges out of the automated behaviour of the system. Although the immediate risk of manipulation by AI is most clearly seen in its use as a tool, the threat of emergent, non-intentional manipulation is clearly relevant and may even be much greater than the risk posed by humans that intentionally use generative AI for manipulative purposes. Hence, the trickery criterion needs to be critically examined.[21]

In summary, the trickery conception faces the biggest challenge in contexts where it generative AI threatens to aggravate existing concerns about manipulation by amplifying the scale of manipulative influence. In lieu of intentions and in lieu of overt features of manipulation, we cannot readily classify emergent forms of manipulation as manipulation on the trickery account.

### The indifference criterion

A proposal that promises to overcome these problems is to identify manipulation with *indifference to some ideal state* rather than some malicious intention to do harm or induce a mistake (Klenk, 2020, 2021c). According to the indifference criterion for manipulation, manipulation is an influence that aims to be effective but is not explained by the aim to reveal reasons to the interlocutor (Klenk, 2021c, 2023).[22]

For example, when a fraudster uses a generative AI application to produce a text message that seems to come from a child in distress to solicit money from a concerned parent, the fraudster's concern will likely be an effective influence (i.e. successful fraud). At the same time, they are indifferent as to *how* they achieve their desired goal. In contrast to the trickery conceptualisation, which interprets the fraudster as intending to trick the victim, the indifference account instead emphasises the fraudster's motivation to use *whichever method works* to reach their goal.[23] Similarly, when generative AI is used to create a political campaign ad that evokes the image of 'foreign' looking people and those images are chosen image *because* they are thought to optimise some desired effect of the campaign (e.g. to ignite people's xenophobia and racial hatred), then that use of the system counts as manipulative (cf. Mills, 1995).

Relatedly, the indifference view can be used to describe manipulation in the behaviour of automated systems. For example, when a recommender system is set to display

---

Footnote 18 (continued)

account, too. Moreover, there are more general considerations about the mistake criterion in terms of false negatives, the criterion might thus be under-inclusive, discussed in Klenk (2021b).

[19] Importantly, the influence is not accidental, since the manipulator did aim to have a particular effect on the target audience. *How* that effect was achieved, however, was unintended.

[20] The result of being manipulated may also come apart from the intention to manipulate, which could be identified independently Klenk (2022b).

[21] The problem is related to but wider than the problem of AI 'hallucinations' where the AI system presents false information as facts. The problem is wider because, as stated above, manipulation cannot be reduced to misleading or false communication. I thank an anonymous referee for pointing out this connection.

[22] Ideas pertinent to the indifference view have also been defended by Gorin (2014b), Mills (1995), and Baron (2014). The account is more systematically developed by Klenk, who first uses the term 'carelessness' (2021), whereas Klenk (2022a, 2022b) introduces the more appropriate term 'indifference' to avoid the misleading impression that manipulation is, overall, lazy or not planned out. Indeed, manipulation is often carefully crafted influence in its aim to be effective, but careless or indifferent only to the aim of revealing reasons to others.

[23] Which may, counterfactually, be a different method than to trick the victim.

content that effectively engages people's attention, and it displays that content for that purpose rather than to reveal reasons to users e.g. about whom to vote for, what to buy, or what to believe, then the recommender system is used manipulatively. Moreover, it might be said that the system itself functions manipulatively (Klenk, 2020, 2022b). This has ramifications for possible future uses of generative AI applications. While current generative AI applications like ChatGPT are not yet capable of fine-tuning their output in pursuit of goals other than text-sequence prediction, attempts to fine-tune generative AI applications with objectives aimed at effective influence are possible future use cases (and already discussed e.g. by Matz et al., 2023). When such future generative AI applications optimise for effective influence on the user (e.g. to increase sales through a customer service application), then their manipulativeness may not come down to anyone's intention (as discussed further below).[24]

The indifference view thus identifies manipulation based on two criteria. First, it only looks at influence that is aimed at a particular goal. In that sense, and in line with most, if not all, of the literature on manipulation, the view excludes influence that is purely accidental from counting as manipulation (see Noggle, 2018).[25] Second, the indifference view then asks why a particular means of influence was chosen to achieve the relevant goal. Manipulative influence is characterised negatively in terms of a choice of a means of influence that is not being explained by the aim to reveal reasons to the target of the influence. The manipulator is, in that sense, "careless" (Klenk, 2021c) or indifferent to revealing reasons to their victims in their choice of the means of influence that they employ. Importantly, the indifference view can be interpreted non-intentionally by thinking about the function of a chosen means of influence. For example, the 'watch next video' choice that a recommender system offers to a user has a particular function, say to induce a target behavior in the user. The indifference view would classify this as manipulation insofar as 'revealing reasons' is not the function of that means of influence.

Notably, the indifference criterion can capture emergent, unwitting manipulation resulting from the sense that generative AI systems act as "stochastic parrots" (Bender et al., 2021). This is one of the chief advantages of the indifference view over the trickery conceptualisation of manipulation. Generative AI systems can be understood as 'bullshitters' in Frankfurt's sense of bullshitting as a type of speech act indifferent to truth (Frankfurt, 2005). Manipulation as a super-category of bullshit (Klenk, 2022a) may not be restricted to malicious intent but more broadly connected to indifference to truth and inquiry.[26] This neatly characterises the 'behaviour' of generative AI systems. They are like "a trickster: they gobble data in astronomical quantities and regurgitate (what looks to us as) information. If we need the "tape" of their information, it is good to pay close attention to how it was produced, why and with what impact" (Floridi, 2023).

However, despite its advantages, the indifference criterion also raises some critical questions. For one thing, the 'ideal state' that manipulators are indifferent to ultimately ideally needs to be specified in more detail to yield a more informative operationalisation. A promising route forward is to investigate what it takes to reveal reasons to interlocutors, as suggested by Klenk (2021b). The vast literature on evidential relations and good deliberation in the philosophical debate promises a suitable starting point. Relatedly, the indifference criterion must be further specified and operationalised to identify manipulation in practice reliably. In particular, what is a reliable sign that indifference explains the choice of the given method of influence? An initial idea is to consider counterfactuals about what method or type of influence would have been chosen if the aim would have been to reveal reasons in a particular situation to a particular user, and to compare the counterfactual output with the system's actual output. A discrepancy could be interpreted as an indicator of *indifference* and, thus, manipulation. Finally, there is a risk that generative AI systems come away as necessarily manipulative (cf. Klenk, 2020), which would not be at all helpful as it blurs the boundary between legitimate and illegitimate uses of those systems (a boundary that plausibly exists).

In summary, the indifference view offers some notable advantages over alternative conceptualisations of manipulation, notably by allowing us to recognise manipulation in situations where intentions are hard, if not impossible, to detect and by avoiding the problems with false positives and false negatives that plague the hidden influence- and bypassing rationality criteria. Unlike disjunctive criteria, it also fares well on the narrow criterion of appropriateness. Like all current conceptualisations of manipulation, however, the indifference criterion relies on further clarification and operationalisation of key terms.

Hence, answering the question of how we can identify manipulation will still require us to answer how we can *identify those criteria in practice*. This is a relevant question because the most plausible criteria for manipulation (the trickery criterion and the indifference criterion) are linked to intentions, purposes or aims, which are not directly observable. This means that designers and regulators need to figure out ways to operationalise the criteria and develop methods to detect them in practice. The design for value approach makes room for that in recommending an iterative process where considerations in the conceptual stage are informed by the empirical- and design stages. Keeping in mind the open question about *how* to choose between a narrow and broad conception of appropriateness, there is room to let design considerations about which conception of manipulation is *implementable* have weight in the choice of conceptualisation.

---

[24] Thanks to an anonymous referee for prompting me to clarify the relevance of future use cases.

[25] Importantly, a goal can but need not be understood in intentional terms. Animals can be said to have goals, as do automated systems, or even simple artefacts based on their use plan, van de Poel (2020), or affordances, Klenk (2021a). In short, goals can be understood in functional terms.

[26] See Klenk (2020) for a discussion of manipulation in relation to bullshit.

The question about picking suitable conceptualisations of manipulation is closely linked to the question of how to ensure that generative AI systems are aligned with the criteria in such a way that they do not generate manipulative content. Following Gabriel (2020), the identification of reliable criteria for manipulation would answer one part of the alignment question. The question about implementation, however, would remain open. Plausibly, as van de Poel (2020) suggests, there would need to be prolonged attention to the system after the initial design stage.

Finally, the ethics of manipulation needs to be evaluated. Though it is generally agreed that manipulation is a morally dubious form of influence, there are open questions about whether or not it is always and categorically morally wrong, or whether manipulation could sometimes be permissible in light of other considerations (cf. Noggle, 2022). Support for the latter position comes from the observation that manipulation is a pervasive part of everyday life, and often not considered to be deeply problematic, as in some marketing or advertising tactics. This would support conceptualising manipulation as pro tanto rather than morally wrong.

Related to this, there is a question about whether and how situational and personal factors may moderate the ethical status of manipulation. Designers and regulators would need to consider if and how those factors have an impact on the moral status of manipulative influence. For example, it may be that the positive impact of some public health communication driven by manipulative generative AI may outweigh the negative value that accrues from the manipulative nature of the influence. Relatedly, some users may willingly adopt personal health assistants that use effective but manipulative influence tactics. In both cases, it is an open question whether these trade-offs are reasonable and ethically legitimate.

**Empirical stage**

One of the core commitments of design for value approaches is the commitment to involve stakeholder perspectives in the design process (Buijsman et al., forthcoming). This usually involves a process of weighing up the conceptualisation of a value developed during the conceptual stage with input gleaned from stakeholders. Looking beyond stakeholders' input toward empirical input more generally, we should consider empirical data that bears on the question of an appropriate conceptualisation of manipulation. At least the following questions specifically related to manipulation need to be addressed in the empirical part of a design for value project:

1. What do relevant stakeholders consider as criteria for manipulation?
2. How should those empirical findings impact conceptual findings?
3. How do stakeholders view the ethical status of manipulation?

The design of non-manipulative generative AI should be informed by empirical findings about criteria for manipulation. But how do people, in fact, distinguish between different forms of influence? The empirical investigation of manipulation is still in its infancy. The studies by Osman and Bechlivanidis are the only ones that explicitly address folk-conceptions of manipulation (Osman, 2020; Osman & Bechlivanidis, 2021, 2022, 2023). An important finding is that judgements about the impact of manipulation and its ethical seriousness differs by context. These findings are valuable starting points. Going forward, it would be interesting to see how the users of a given generative AI application think about manipulation in the aim of design approaches to consider especially the views of stakeholders in the design process. An important question is whether and how the views of different groups of stakeholders differ regarding criteria for manipulation. For instance, are there political or personal factors that moderate how people distinguish manipulative from non-manipulative influence? Next to quantitative research paradigms familiar from the social sciences, researchers and regulators can draw on established methods for design for values approaches, such as focus groups or participatory design, to address these questions.

More generally, there is a need for further studies of the folk concept of manipulation. While such findings do not settle which conceptualisations of manipulation are appropriate, they will serve as valuable reference points. Are there (aspects of) folk conceptualisations not covered by any of the current theoretical conceptualisations? Which theoretical conceptualisations (most closely) match the ordinary conceptualisation of manipulation? What factors influence how people think about manipulation? Are there personal or situational factors that influence whether or not people make reliable judgements about manipulation and its (dis-)value? Answering these questions is relevant both from a narrow and broad criterion for appropriateness, since the answers may bear on the accuracy of the conceptualisation or its moral appropriateness.

Gathering empirical insights into judgements about manipulation will raise the question of how those findings should be combined with the conceptual findings of the previous step. Should empirical findings lead researchers to revise reliable criteria for manipulation? If so, to what extent? Presumably, manipulation is a phenomenon that is socially constructed in the limited sense that it depends on people and social structures to exist (Hacking, 1999), but it is an open question whether its criteria are entirely depended on what people think of it. There are, however, strong reasons to suspect that the criteria for manipulation are not entirely up for grabs: they are not entirely determined by what people think of manipulation. To illustrate, consider the case of an generative AI system that manages to influence users' views on what counts as manipulation. Users may then judge that bypassing their reason, influencing them covertly, and trying to induce mistakes in them is a legitimate form of persuasion, rather than manipulation. Designers and regulators should not take that result to revise their theory of manipulation entirely. How, precisely, the revision should work, however is an intricate, and open question. The literature on the significance of empirical, experimental philosophy may offer relevant pointers on this question (Knobe & Nichols, 2017), as well as the literature on applying theories in the context of bioethics (Beauchamp & Childress, 2019).

Quite independently of findings about conceptualisations of manipulation, empirical findings can help us understand more about the value of different conceptualisations. For example, empirical findings are clearly relevant in determining the impact of manipulation. Though manipulation seems like a prima facie problematic type of influence, as discussed in the previous section, it matters for the ethical assessment whether it has particularly pernicious consequences. So far, however, we know very little about the impact of manipulation. There is a widespread assumption that manipulation is antithetical to autonomy (Susser et al., 2019b), but that view has yet to be corroborated from an empirical perspective (Klenk & Hancock, 2019), and we already know that people's judgements about manipulation's impact are more nuanced (Osman & Bechlivanidis, 2021). We also know that generative AI can have an impact on people's moral judgements (Krügel et al., 2023), but it is unclear whether and why the influence in question qualifies as manipulation or not.

Finally, considerations about the need to consult stakeholders and to reflect on how their views should impact criteria for manipulation will also apply to the empirical investigation of the ethics of manipulation. Can manipulation be 'made' permissible insofar as people consent to it? Do people consent to manipulation? If so, under which circumstances and in what contexts? An important question here is how to align empirical findings which suggest more lenient takes on the ethics of manipulation with the strong regulatory aversion against manipulative influence that already applies to generative AI.

## Design stage

Insights from the conceptual and the empirical stage will eventually have to be translated and transformed into concrete design requirements for generative AI applications. Given the present focus on appropriate conceptualisations of manipulation, these questions are out of scope for this paper. Nonetheless, at least two broad questions that crop up at the design stage have a bearing on appropriate conceptualisations nonetheless.

For one, it might be thought that the conceptualisation question could be settled by design. Specifically, there may be ways to address the conceptualisation question through different alignment approaches in AI, which ultimately bottom-out in a machine learning approach.[27] Kenton et al. (2021) discuss as the option to include human preferences in the training of the generative AI system (Christiano, 2017). Roughly, this means that the output of the system is fine-tuned in light of user feedback. For example, human souls classify sample outputs of the system as (more or less) manipulative to fine-tune the system with this feedback. More precisely, Ouyang et al. (2022) describe a process to alignment that fine-tunes the output of a Large Language Model in light of human-generated output (which is used in a supervised learning model) and human rankings of system-generated output (to train a reward model, which then fine-tunes the supervised baseline using reinforcement learning). Ouyang et al. (2022) demonstrate that the resulting model shows improvements over the outputs of GPT-3 (which are not fine-tuned by the described process) in several ethically significant domains like the toxicity and truthfulness of the output.

Naturally, such an approach depends on the ability of human labelers to spot manipulation, and it raises questions about combining users' perspectives with theoretical insights discussed in the previous section. Kenton et al. (2021) do not discuss that users have a questionable track-record of discovering manipulation. Ouyang et al. (2022) are sensitive to the issue that the success of their alignment approach depends on the quality of the feedback provided by human labellers, and they suggest that the quality of human feedback may depend on a variety of personal and situational factors. In this context, the *philosophical* and '*folk-psychological*' disagreement about conceptualisations of manipulation must be stressed more. At least, the current empirical findings on manipulation suggest human judgements about manipulation differ, sometimes quite strikingly, from the conceptualisations defended in the philosophical literature. Therefore, an important question in the design stage concerns not just the technical implementation of human feedback for fine-tuning, but the empirical- and theoretical investigation of the reliability of user judgements in learning from human feedback.

---
[27] Thanks to an anonymous referee for suggesting this formulation, and for prompting me to clarify this point.

There is an important sense in which design requirements can only be judged for their appropriateness after testing and experimenting. An early chatbot, TAI by Microsoft, illustrated how an initially well-functioning system got off the rails over time by updating its behaviour in response to user feedback. So, manipulation in generative AI may only arise after some time, after deployment, and design needs to take measures to deal with that risk. For that reason, van de Poel (2020) advocates for prolonged monitoring, in addition to considerations about the appropriately aligning the model.

Moreover, the broad criterion of appropriateness may allow us to consider pragmatic considerations as relevant for choosing a conceptualisation of manipulation. On the one hand, considerations about the technology's capability may prompt us to adjust the conceptualisation of manipulation. Consider that most current cases of manipulation with generative AI involve a human in the loop that uses generative AI to generate manipulative content (see Goldstein et al., 2023). In such cases, design against manipulation may rely on a conceptualisation of manipulation that refers to intentions, since we can ask about the intentions of the human in the loop. But this may change. While the leading current generative AI applications produce output that essentially predicts the next text token on a webpage from the internet (Ouyang et al., 2022), future applications may fine-tune that output with e.g. the aim to improve persuasiveness, possibly by incorporating information about personal attributes of the human user. Matz et al. (2023) already demonstrate that GPT-3 can be prompted to produce personalised and more persuasive outputs when the 'human prompter' is able to match prompt and target.[28] Future applications will likely attempt to automate the process of obtaining persuasion profiles of targets and producing persuasive prompts using generative AI, thus removing the human from the loop. Given the aim to design for non-manipulation also in cases like this, and doubts about the intentionality of generative AI, there is reason to favour a conceptualisation that makes no reference to intention. Pepp et al. (2022) already discuss this option in some detail.[29] In this way, concrete considerations about the practical use of generative AI applications that crop up at the design stage will might have a bearing on the conceptual stage of the design process.

On the other hand, there may be moral reasons to pick a conceptualisation that is applicable to the technology if and insofar as such a conceptualisation serves a worthy moral goal. Calling some applications that use generative AI *manipulative* may, for instance, lead to desirable consequences because they come under public scrutiny or in the scope of regulation All this depends, of course, on the appropriateness of the broad criterion in the first place.

In summary, even though questions at the design stage are not of primary concern for picking a conceptualisation of manipulation, technical and moral aspects may yet make the design stage relevant, given a broad criterion of appropriateness for choosing conceptualisations.

# Conclusion

Generative AI brings enormous promise and peril. It may enable effective, automated influence at scale. This can be used for good, for instance in meaningful and ethical communication or in the design of digital health assistants. But it also harbours the risks of manipulation. This article introduced a research agenda focused on designing generative AI systems for non-manipulation to make good on its promise and avoid the peril. It demonstrated that if we want to design for non-manipulation, which everyone interested in responsible and trustworthy AI should be concerned with, we must begin with an appropriate conceptualisation of the phenomenon.

Apart from drawing attention to pertinent research questions concerning manipulation and generative AI, the main upshot of the article is a reasonable, brief overview of not only the importance of choosing the right conceptualisation of manipulation but also some of the key considerations for doing so. Clearly, both the general point about how to choose appropriate conceptualisations, and the specific points about different possible ways to conceptualise manipulation are mere beginnings. Key questions such as 'how should we pick conceptualisation?' and intricate points from the debate about manipulation remain beyond the scope of this paper. In particular, questions about the value of a given conceptualisation vis-à-vis its practical implementability in current alignment approaches are crucial, and difficult to answer.

Each dimension—conceptual, empirical, and design—should, in future work, be further elaborated on to outline the research questions in more detail and to critically consider different attempts at answering them. Given the aim to direct the debate in a fruitful direction, however, these omissions are deemed justified. In light of existing legal and regulatory measures against manipulation, and moral concerns that are aggravated by generative AI, the questions outlined in this article should contribute to some progress toward the responsible innovation of generative AI.

---

[28] To wit, the human prompter is able to assess the persuasion profile of the target, subsequently prompt GPT-3 to produce an output that matches that profile, and then present the target with that output.

[29] The emerging work on conceptual engineering in the ethics of technology offers further examples concerning, e.g., notion of responsibility Veluwenkamp and van den Hoven (2023), Himmelreich and Köhler (2022).

**Acknowledgements** I am thankful to Caroline Figueiredo, and to two anonymous referees for helpful feedback.

**Author contributions** N/A (single author).

**Funding** The author's work on this paper has been part of the project Ethics of Socially Disruptive Technologies that has received funding from the Dutch Organisation of Scientific Research.

**Data and materials availability** All data is available in the MS.

## Declarations

**Conflict of interest** No competing interests.

**Ethical approval and consent to participate** N/A.

**Consent for publication** Consent for publication is given.